\documentclass[prb,twocolumn,superscriptaddress,showpacs,floatfix]{revtex4}
\usepackage{amssymb}
\usepackage{amsfonts}
\usepackage{subfigure}
\usepackage{graphicx}
\usepackage{dcolumn}
\usepackage{bm}
\usepackage{hyperref}
\usepackage[mathlines]{lineno}
\usepackage{amsmath}

\preprint{}

\begin{document}

\title{Formation of Molecular-Orbital Bands in a Twisted Hubbard Tube: Implications for Unconventional Superconductivity
in K$_2$Cr$_3$As$_3$}

\author{Hanting Zhong}
\affiliation{Condensed Matter Group,
  Department of Physics, Hangzhou Normal University, Hangzhou 310036, China}

\author{Xiao-Yong Feng}
\email{fxyong@163.com}
\affiliation{Condensed Matter Group,
  Department of Physics, Hangzhou Normal University, Hangzhou 310036, China}
\affiliation{Hangzhou Key Laboratory of Quantum Matter, Hangzhou
Normal University, Hangzhou 310036, China}

\author{Hua Chen}
\affiliation{International Center for Quantum Materials and School
of Physics, Peking University, Beijing 100871, China}
\affiliation{Collaborative Innovation Center of Quantum Matter,
Beijing 100871, China}

\author{Jianhui Dai}
\email{daijh@zju.edu.cn}
\affiliation{Condensed Matter Group,
  Department of Physics, Hangzhou Normal University, Hangzhou 310036, China}
\affiliation{Hangzhou Key Laboratory of Quantum Matter, Hangzhou
Normal University, Hangzhou 310036, China}


\begin{abstract}
We study a twisted Hubbard tube modeling the [CrAs]$_{\infty}$
structure of quasi-one-dimensional superconductors $A_2$Cr$_3$As$_3$
($A=$ K, Rb, Cs). The molecular-orbital bands emerging from the
quasi-degenerate atomic orbitals are exactly solved. An effective
Hamiltonian is derived for a region where three partially filled
bands intersect the Fermi energy. The deduced local interactions
among these active bands show a significant reduction compared to
the original atomic interactions. The resulting three-channel
Luttinger liquid shows various interaction-induced instabilities
including two kinds of spin-triplet superconducting instabilities
due to gapless spin excitations, with one of them being superseded
by the spin-density-wave phase in the intermediate Hund's coupling
regime. The implications of these results for the alkali chromium
arsenides are discussed.
\end{abstract}

 \maketitle

\begin{figure}
\includegraphics[width=8.0cm]{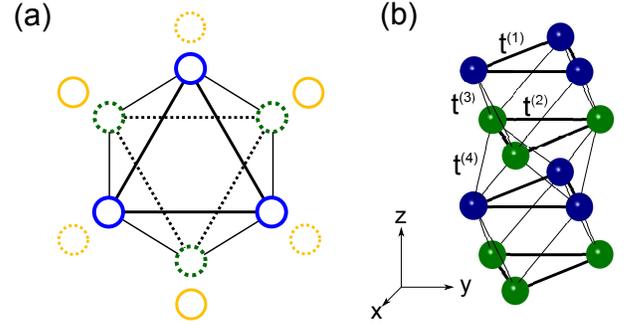}
\caption{(Color online) (a) A CrAs cluster in the $ab$ plane. The
solid (dotted) circles connected by the solid (dotted) lines
represent the Cr atoms in the first (second) triangle in a unit
cell. The isolated outer solid (dotted) circles represent the As
atoms in the corresponding planes. (b) A Q1D CrAs tube. The blue
(green) filled circles represent the Cr atoms in each triangles. The
As atoms are not shown. } \label{CrAstube}
\end{figure}

{\it Introduction.}---Recently, the alkali chromium arsenides
$A_2$Cr$_3$As$_3$ ($A=$ K,Rb,Cs) have been found as a new family of
inorganic quasi-one-dimensional (Q1D) superconductors with strong
electron correlations.\cite{Bao14,Tang14,Tang15} The basic building
block of these compounds is the [CrAs]$_6$ cluster consisting of two
conjugated triangular complexes [CrAs]$_3$ as shown schematically in
Fig.\ref{CrAstube}(a). They are aligned along the c axis forming a
[CrAs]$_\infty$ tube, and intercalated by $A^{+}$ cations forming a
hexagonal lattice. The density functional theory (DFT)
calculations\cite{Jiang14,Wu15} predict a three-dimensional (3D)
Fermi surface (FS) sheet ($\gamma$ band) and two Q1D FS sheets
($\alpha$ and $\beta$ bands), essentially due to the Cr $3d$
electrons. The NMR experiment\cite{Zhi15} has revealed a power law
behavior of the spin-lattice relaxation rate, manifesting the
Luttinger liquid feature above $T_c$. The penetration depth
measurement\cite{Pang15} has evidenced a line nodal feature in the
pairing state below $T_c$.

Because of the existing 3D $\gamma$ band, whether the
superconductivity is solely originated from the Q1D structure of
$A_2$Cr$_3$As$_3$ is uncertain. In fact the nearly isotropic 3D bulk
CrAs compound shows superconductivity with $T_c \sim $ 2.2 K under
pressure of $\sim 0.7$ GPa.\cite{Wei14,Kotegawa14} Interestingly,
Zhou {\it et al.} pointed out that an $f$-wave pairing state could
arise from the 3D band with a node line while a fully gapped
$p$-wave pairing state could dominate at the Q1D band.\cite{Zhou15}
Such triplet superconductivity, with some variations in spatial
symmetry\cite{Wu15b}, could be driven by ferromagnetic fluctuations
within the sublattice of Cr atoms\cite{Jiang14,Wu15b}.

In order to understand the formation of the low energy bands, it is
particularly important to understand the electronic property of {\it
a single fundamental }[CrAs]$_{\infty}$ {\it tube}. In this Letter,
we model this system by a twisted Hubbard lattice composed of
triangular complexes coupled along the c axis with the glide
reflection symmetry as shown in Fig.\ref{CrAstube}(b).\cite{note1}
In each unit cell there are six Cr atoms, each with five $3d$ atomic
orbitals (AOs). The influence of the As $4p$ orbitals can be
effectively accounted for the indirect hopping of Cr-$3d$ electrons.
So the model involves thirty energy bands in total. In the realistic
case, fortunately, only three partially filled bands are active in
the low energy regime. We will explicitly show how these bands come
from the molecular orbitals (MOs) of
[CrAs]$_{6}$\cite{Jiang14,Zhou15}. Our purpose is then to understand
their cooperative low temperature physics within the Luttinger
liquid approach. The proposed effective model is of interest in its
own right as we shall explore in the following.

{\it Model Hamiltonian.}---The Hubbard model for a single
[CrAs]$_{\infty}$ tube is expressed as $H=H_0+H_{\rm int}$, where,
$H_0$ represents the noninteracting part consisting of the
tight-binding kinetic energy and the crystalline electric field
(CEF) splitting,
\begin{eqnarray}
H_0&=&-\sum_{{\bf r}{\bf r'}mm'\sigma}t^{({\bf r},{\bf
r}')}_{mm'}d^{\dagger(m)}_{{\bf r}\sigma} d^{(m')}_{{\bf r'}\sigma}
+\sum_{{\bf r}m\sigma}E_{{\bf r}m}{n}^{(m)}_{{\bf r}\sigma}.
\label{h0}
\end{eqnarray}
Here, $d^{(m)}_{{\bf r}\sigma}$ denotes the annihilation operator of
Cr $3d$ electrons at the site ${\bf r}$ with angular momentum $m
=0,\pm 1,\pm 2$, spin $\sigma =\uparrow,\downarrow$. $n^{(m)}_{{\bf
r}\sigma}$ and ${\bf{S}}^{(m)}_{{\bf r}}$ are the corresponding
density and spin operators. The two twisted Cr triangles could have
different $E_{{\bf r}m}=E^{(1)}_{m}$, $E^{(2)}_{m}$, accountable for
the possible occupation difference\cite{Wu15}, while
\begin{eqnarray} H_{\rm int}&=&U\sum_{{\bf
r}m}n^{(m)}_{{\bf r}\uparrow} n^{(m)}_{{\bf
r}\downarrow}+\frac{2U'-J_H}{4}\sum_{{\bf r}m\neq m'\sigma\sigma'}
n^{(m)}_{{\bf
r}\sigma}n^{(m')}_{{\bf r}\sigma'} \nonumber\\
&& -~J_H \sum_{{\bf r}m\ne m'}{\bf{S}}^{(m)}_{{\bf r}}
\cdot{\bf{S}}^{(m')}_{{\bf r}}\\
&& +~J_p \sum_{{\bf r}m\ne m'}d^{\dagger(m)}_{{\bf
r}\uparrow}d^{\dagger(m)}_{{\bf r}\downarrow} d^{(m')}_{{\bf
r}\downarrow}d^{(m')}_{{\bf r}\uparrow}\nonumber \label{hint}
\end{eqnarray}
represents the local interactions including the intraorbital Coulomb
interaction $U$, the interorbital Coulomb interaction $U'$, the
Hund's coupling $J_H$, and the pair-hopping $J_p$, respectively.

There are four kinds of adjacent intraorbital hoppings
$t^{(i)}_{mm}\equiv t^{(i)}_{m}$ ($i=1-4$), corresponding to the
nearest-neighbor (NN) sites in the first and second triangles, and
those between the intracell and intercell triangles, respectively,
as illustrated in Fig.\ref{CrAstube}(b). Because of the metallic
bonding among Cr atoms, the direct orbital mixings are relatively
small, and the indirect hybridization is mainly bridged by the As
$4p$ orbitals. So it is legitimate to consider a simpler situation
for the adjacent interorbital hopping: $t^{(i)}_{mm'}=\eta
t^{(i)}_m\delta_{|m|,|m'|}$ for $m\neq m'$, with $|\eta|<1$. In this
situation, the atomic orbitals are quasidegenerate as the
nonvanishing mix terms are isotropic in space\cite{note2}. Finally,
we include the next NN intraorbital hopping $t^{(5)}_{m}$ along the
tube direction.

{\it Molecular-orbital bands.}---Denoting each site by ${\bf
r}=(n,a,\xi)$, with $a=1,2,3$ being the location in the first
($\xi=1$) or second $(\xi=2)$ triangles in the $n$th unit cell, it
is convenient to introduce a base $d^{(m)}_{n}=\left
(d^{(m)}_{(n,1,1)},d^{(m)}_{(n,1,2)},d^{(m)}_{(n,2,1)},d^{(m)}_{(n,2,2)},
d^{(m)}_{(n,3,1)},d^{(m)}_{(n,3,2)} \right)^T$  for the atomic
$m$-orbital in the $n$th unit cell (the spin index $\sigma$ is
implied). For $m=0$, this base accommodates a representation for the
$C_3$ rotational symmetry, leading to six MOs corresponding to $\it
E$,$\it E'$,$\it A$, and $\it A'$ states, respectively\cite{Tao}.
For $m=\pm1$ or $\pm 2$, we need to introduce a set of new base
${\tilde d}^{(\pm|m|)}_{n}=\frac{1}{\sqrt 2}[d^{(m)}_{n}\pm
d^{(-m)}_{n}]$. Thus for a single [CrAs]$_6$ cluster, we have thirty
MOs defined by $C^{(\tau)}_{n}=( {\hat R}\otimes{\hat Q}_0){\tilde
d}^{(\tau)}_{n}$ for $\tau=0$ (denoting ${\tilde d}^{(0)}_n\equiv
d^{(0)}_n$), $\pm 1$ and $\pm 2$, respectively, with
($\omega=e^{i\varphi}$, $\varphi=2\pi/3$)
\begin{eqnarray}
{\hat R}&=& \frac{1}{\sqrt 3} \left(
\begin{array}{ccc}
 1 & 1 & 1 \\
 1 & \omega & \omega^{-1} \\
 1 & \omega^{-1}& \omega
\end{array}
\right),
{\hat Q}_0= \frac{1}{\sqrt 2}
\left(
\begin{array}{cc}
 1 &  -1 \\
 1 & 1
\end{array}
\right).~~~ \label{matrices}\end{eqnarray} Note that the eigenstates
of ${\hat R}$ with eigenvalues $\lambda_1=2$ and
$\lambda_2=\lambda_3=-1$ constitute of representations $A$ and $E$
(or $A'$ and $E'$), respectively.

When the triangles are coupled along the c axis via the intercell
hopping $t^{(4)}_{mm'}$, we can extend ${\hat Q}_0$ to the momentum
$k$-resolved matrix ${\hat Q}^{(\tau)}_{a}(k)$ so that the Bloch
form $C^{(\tau)}_{k}=\left
(c^{(\tau)}_{(k,1,1)},c^{(\tau)}_{(k,1,2)},c^{(\tau)}_{(k,2,1)},c^{(\tau)}_{(k,2,2)
},c^{(\tau)}_{(k,3,1)},c^{(\tau)}_{(k,3,2)} \right)^T$ is still a
natural base diagonalizing $H_0$, leading to thirty MO bands labeled
by the eigenenergies ${\cal E}^{(\tau)}_{(a,\xi) }(k)$. Here the
subscript $\xi=1,2$ corresponds to the antibonding or bonding bands,
respectively, due to the twisted structure. The explicit expressions
of ${\hat Q}^{(\tau)}_{a}(k)$ and ${\cal E}^{(\tau)}_{(a,\xi) }(k)$,
which also depend on the orbitals $\tau (=0,\pm 1,\pm 2)$ and $C_3$
eigenvalues $\lambda_a (a=1,2,3)$, are presented in the Supplemental
Material (SM)\cite{SM}. A set of subscripts $(a,\xi)$ determines the
symmetry property of the corresponding MO bands.

\begin{figure}
\includegraphics[width=8.0cm]{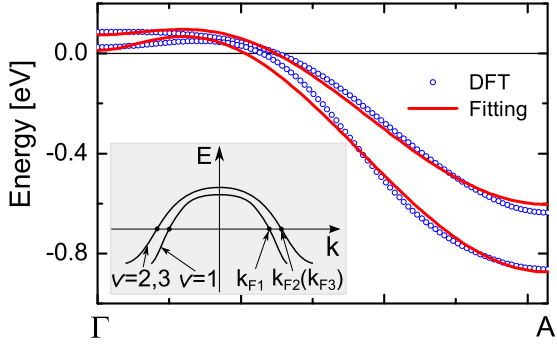}
 \caption{(Color online) Fitting the band structure: The upper band is
twofold degenerate. The lower inset is the schematic picture for the
three partially filled bands with three pairs of Fermi points. }
\label{fit}\end{figure}

We fitted the DFT band structure along the tube direction using the
obtained MO bands within $\tau=0,\pm 2$, while the bands with
$\tau=\pm 1$ are fairly away from the Fermi energy as revealed in
the DFT calculations\cite{Jiang14,Wu15}. The three partially filled
DFT bands, i.e., the 3D $\gamma$ band characterized mainly by the
$d_{z^2}$ orbital ($m=0$), the Q1D $\alpha$ and $\beta$ bands
characterized mainly by the $d_{xy}$ and $d_{x^2-y^2}$ orbitals (
$|m|=2$), are all holelike near the $\Gamma$ point ($k=0$) and
electronlike near the $A$ point($k=\pi$). Therefore, the $\gamma$
band corresponds to the singlet MO band labeled by $(\tau
=0,a=1,\xi=1)$. The $\alpha$ and $\beta$ bands, which are degenerate
along the whole $\Gamma\rightarrow A$ direction, correspond to the
doublet MO bands labeled by $(\tau =-2,a=2,\xi=2)$ and $(\tau
=-2,a=3,\xi=2)$, respectively.  The best fitting using ${\cal
E}^{(0)}_{(1,1)}$, ${\cal E}^{(-2)}_{(2,2)}={\cal E}^{(-2)}_{(3,2)}$
is shown in Fig.\ref{fit} \cite{SM}. Here, the tight-binding
parameters are not uniquely determined because the number of these
parameters exceeds eight necessary coefficients in the fitting. On
the other hand, the precise values of the fitting parameters are not
important in the present study. As we shall find later, only
symmetry property of the MO bands and local interactions between
them play a crucial role in the Luttinger liquid approach.

For simplicity, from now on, we shall use the band subscript $\nu$
($=1,2,3$) to account for the three MO bands intersecting the Fermi
energy. These active MO bands are associated with phases
$\varphi_\nu=0,2\pi/3,-2\pi/3$, or chiralities $\vartheta_{\nu}=0$,
$1$, and $-1$, respectively. In the full 1D Brillouin zone, there
are three pairs of Fermi points ($k_{F_{\nu}}, -k_{F_{\nu}}$),
satisfying $0<k_{F_{1}}<k_{F_{2}}=k_{F_{3}}<\pi$, as schematically
shown in the inset of Fig.\ref{fit}. By integrating out all inactive
bands, we obtain the effective theory describing the low energy
property of the active bands:
\begin{eqnarray} H_{\rm eff}=\sum_{k\nu_i\sigma}{\cal
E}_{\nu}(k){\hat n}_{k\nu\sigma}+\sum_{n}H^{(n)}_{\rm int}.
\label{heff}
\end{eqnarray}
Where, ${\hat n}_{k\nu\sigma}=c^{\dagger}_{k\nu\sigma}
c_{k\nu\sigma}$ is the density operator of electrons in the $\nu$th
MO band, $H^{(n)}_{\rm int}$ the residual short-range interactions
in the MOs in the $n$th unit cell, given by
\begin{eqnarray}
H^{(n)}_{\rm int}&=&\sum_{\nu} {\tilde U}_{\nu} {\hat
n}_{\nu\uparrow}(n){\hat n}_{\nu\downarrow}(n)
+\sum_{\nu\neq\nu'\sigma\sigma'} {\tilde U}_{\nu\nu'} {\hat
n}_{\nu\sigma}(n){\hat n}_{\nu'\sigma'}(n)\nonumber\\
&-&\sum_{\nu\neq\nu'}{\tilde J}_{\nu\nu'} {\hat {\bf
S}}_{\nu}(n)\cdot{\hat {\bf S}}_{\nu'}(n)\\
&+&{\tilde J}_{123}\left[
c^{\dagger}_{1\uparrow}(n)c^{\dagger}_{1\downarrow}(n)
c_{2\downarrow}(n)c_{3\uparrow}(n)+(2\leftrightarrow
3)+H.c.\right].\nonumber \label{heffint}\end{eqnarray} In this
expression, the electron annihilation operator in the $n$-th unit
cell is defined by the Fourier transformation $
c_{\nu\sigma}(n)=\frac{1}{\sqrt{2\pi}}\sum_{k}e^{ik n
c_0}c_{k\nu\sigma}$ (with $c_0$ the lattice spacing taken as unit).
Only those terms preserving the neutrality condition
$\sum_{\nu_i}\vartheta_{\nu_i}=0$ could survive. The matrix ${\hat
Q}_0$ is used in deducing Eq.(5) as the short-range interactions are
mainly due to the slowly varying part, leading to ${\tilde
U}_{1}=U/6$, ${\tilde U}_{2}={\tilde U}_{3}=(U+U'+J_H+J_p)/12$;
${\tilde U}_{12}={\tilde U}_{13}=U'/12-J_H/24$; ${\tilde
U}_{23}=(U+U'+J_H+J_p)/48$; ${\tilde J}_{12}={\tilde J}_{13}=J_H/6$;
${\tilde J}_{23}=(U+U'+J_H+J_p)/12$; and ${\tilde J}_{123}=J_p/6$.
The influence of inactive bands is mainly accounted to the
renormalized tight-binding parameters.

{\it The Luttinger liquid in the weak-coupling regime.}---We now
take the continuous limit, linearize the active bands near the Fermi
points, and decompose the electron operator into right and left
moving components like $ c_{\nu\sigma}(z)\approx
e^{-ik_{F_{\nu}}z-i\varphi_\nu}L_{\nu\sigma}(z)
+e^{+ik_{F_{\nu}}z+i\varphi_\nu}R_{\nu\sigma}(z)$. Here, $z=n c_0$
is the spatial coordinate along the tube direction, $R_{\nu\sigma}$
and $L_{\nu\sigma}$ represent the right and left moving fermions
describing the low energy excitations near the Fermi points
($k_{F_{\nu}}$, $-k_{F_{\nu}}$) with linear dispersion $\pm
v_{F_{\nu}} k$. The long-wavelength, low-energy effective
Hamiltonian (density) is given by $ {\cal H}_{\rm eff}={\cal
H}_{0}+{\cal H}_{\rm int}$, where ${\cal
H}_{0}=\sum_{\nu,\sigma}(iv_{F_{\nu}})
[R^{\dagger}_{\nu\sigma}\partial_{z}R_{\nu\sigma}-L^{\dagger}_{\nu\sigma}\partial_{z}L_{\nu\sigma}]$
is the kinetic part, and ${\cal H}_{\rm int}$ includes various
residual interactions which are usually expressed in terms of the
$g$-ology\cite{Solyom79,Giamarchi04}. We shall assume the Fermi
velocities $v_{F_{\nu}}$ to be the same as this does not influence
the nature of superconductivity we concern. The corresponding
one-loop renormalization group (RG) equations resemble those for
three-leg Hubbard ladders\cite{Arrigoni96,Lin98} or a variant of
carbon nanotubes\cite{Krotov97,Gonzalez05,Carpentier06}. The
instabilities of these RG equations are classified routinely: (i)
the intraband instabilities as those developed in the single-channel
Luttinger liquid\cite{note3}, and (ii) the interband instabilities
as those developed in the two-channel band Luttinger liquid. Note
that the three-band interaction in Eq.(5) does not lead to the
peculiar three-band instability suggested in Ref.\cite{Carpentier06}
as shown in the SM\cite{SM,note4}. All these suggest the validity of
the conventional bosonization approach based on spin-charge
separation, where various ordering instabilities can be determined
by Luttinger parameters. The new ingredients here are the peculiar
symmetry surviving in the active MO bands and their dependence on
local electron interactions.

The right- and left-moving fields are then expressed in terms of the
charge fields ($\phi_{\nu,c}$, $\theta_{\nu,c}$) and the spin fields
($\phi_{\nu,s}$, $\theta_{\nu,s}$) (for each $\nu=1,2,3$) by
\begin{eqnarray}
R_{\nu,\sigma}(z)=\frac{F_{R,\nu\sigma}}{\sqrt{2\pi
c_0}}e^{i\sqrt{\pi/2}
(\theta_{c,\nu}+\sigma\theta_{s,\nu}-\phi_{c,\nu}-\sigma\phi_{s,\nu})},\nonumber\\
L_{\nu,\sigma}(z)=\frac{F_{L,\nu\sigma}}{\sqrt{2\pi
c_0}}e^{i\sqrt{\pi/2}
(\theta_{c,\nu}+\sigma\theta_{s,\nu}+\phi_{c,\nu}+\sigma\phi_{s,\nu})}.
\label{leftright}
\end{eqnarray}
The Klein factors $F_{R,\nu\sigma}$ and $F_{L,\nu\sigma}$ ensure the
fermionic statistics between the right and left moving fermions.
Next, in order to diagonalize the kinetic part, we need to introduce
a set of new base
\begin{eqnarray}
\tilde{\phi}_{\gamma,i}&=&\eta_{\gamma,i}\left(q_{\gamma, i}\phi_{\gamma ,1}+\phi_{\gamma ,2}+\phi_{\gamma,3}\right),\nonumber\\
\tilde{\phi}_{\gamma,3}&=&\frac{1}{\sqrt{2}}\left(-\phi_{\gamma,2}+\phi_{\gamma,3}\right),
\label{newbase}\end{eqnarray} where $\gamma=s,c$, $q_{\gamma,
i}=-\frac{b_{\gamma}+(-1)^{i}\sqrt{8a_{\gamma}^2+b_{\gamma}^2}}{2a_{\gamma}}$
for $i=1,2$, $a_c= \frac{2\tilde{U}_{12}}{\pi}$, $b_c =
\frac{2\tilde{U}_{23}}{\pi}$, $a_s= -\frac{\tilde{J}_{12}}{2\pi}$,
$b_s = -\frac{\tilde{J}_{23}}{2\pi}$, $\eta_{\gamma,i}$ are the
normalization constants. Similar relationships apply to the fields
$\theta_{\gamma,i}$ and $\tilde{\theta}_{\gamma,i}$ for $i=1,2,3$.
Hence, we arrive at the following three-channel Tomonaga-Luttinger
liquid Hamiltonian:
\begin{eqnarray}
\tilde{H}_0=\int dz\sum_{i=1,2,3,\gamma=s,c}
\left[\frac{v_F}{2}(\nabla\tilde{\theta}_{\gamma,i})^{2}+\lambda_{\gamma,i}(\nabla\tilde{\phi}_{\gamma,i})^{2}\right]
\label{hluttinger}\end{eqnarray} where,
$\lambda_{\gamma,i}=t_{\gamma}+\frac{1}{2}\left[b_{\gamma}-(-1)^{i}\sqrt{8a_{\gamma}^2+b_{\gamma}^2}\right]$
for $i=1,2$, and $\lambda_{\gamma,3}=t_{\gamma}-b_{\gamma}$,
$t_c=\frac{v_F}{2}+\frac{\tilde{U}_1 }{2\pi}$,
$t_s=\frac{v_F}{2}-\frac{\tilde{U}_1 }{2\pi}$. Therefore, the
Luttinger parameters are obtained explicitly by
\begin{eqnarray*}
\begin{array}{l}
K_{c,i}=\left[1+\frac{U}{4\pi v_F}-\frac{(-1)^{i}}{12\pi v_F}\sqrt{8(2U-5J_H)^2+U^2}\right]^{-\frac{1}{2}},\\
K_{s,i}=\left[1-\frac{U}{4\pi v_F}-\frac{(-1)^{i}}{12\pi
v_F}\sqrt{8J^2_H+U^2}\right]^{-\frac{1}{2}}
\end{array}
\end{eqnarray*}
for the channels $i=1,2$, respectively, and $
 K_{c,3}=K_{s,3}=1 $ for the third channel $i=3$. Here, we have adopted the conventional relations $J_p=J_H$ and
$U'=U-2J_H$, reflecting the rotational symmetry of the original
AOs\cite{Castellani78}.

\begin{figure}
\includegraphics[width=5.5cm]{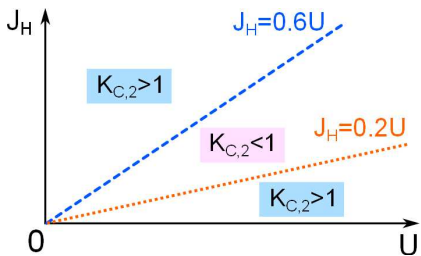}
\caption{(Color online) Luttinger parameters for the channel-2:
$K_{s,2}>1$ everywhere, and $K_{c,2}<1$ only in the intermediate
regime between the lines $J_H=0.2U$ and $J_H=0.6U$.} \label{phase2}
\end{figure}

Now since $U>0$ and $J_H>0$, one can find that: (i) $K_{c,1}<1$ in
the entire region and $K_{s,1}<1$ only when $U<J_H$; (ii)
$K_{c,2}<1$ in the region $0.2U<J_H<0.6U$ and $K_{s,2}>1$ in the
entire region. Specifically, in the physically relevant regime,
$U>J_H$, the spin excitations are always gapless, so $K_{s,i}$ could
be fixed to the unit due to the spin-SU(2) symmetry. Because
$K_{c,1}<1$, the channel-"1" is in the spin-density-wave (SDW)
phase\cite{Solyom79,Giamarchi04}. The channel-"3" involves the
antibonding of the MO bands $\nu=2,3$ as shown in
Eq.(\ref{newbase}). In this channel both spin and charge excitations
are critical. Because of the absence of a spin gap, the dominating
superconducting instability is the interband spin triplet
pairing\cite{Solyom79,Giamarchi04}, driven by the interband
scattering between the two Q1D $\alpha$ and $\beta$ bands. The
intriguing case is the channel-"2", whose property depends on the
ratio $J_H/U$.  We plot in Fig. \ref{phase2} the phase diagram
determined by the Luttinger parameters in this channel.
We find that $K_{c,2}>1$ in the regimes separated by the
orange-dotted and blue-dashed lines, respectively. In these two
separated regimes, the dominating instability is still the
spin-triplet pairing\cite{Solyom79,Giamarchi04}. But in the
intermediate regime where $K_{c,2}<1$, the SDW instability
dominates. It should be noticed that in either case where the
interband triplet superconducting instabilities dominate, the
spin-singlet superconducting instability is the subdominating
instability\cite{Giamarchi04,note5}.

In order to see whether the above Luttinger liquid results are
robust against deviations from the atomic orbital rotational
symmetry, we have also considered a small deviation $\Delta U$ away
from the rotational symmetry by assuming $U'=U+\Delta U-2J_H$. As
shown in the SM, the channel-3 is still in the critical phase and
the role of $\Delta U$ is to modify the value of U in a simple
manner so that the results remain unchanged\cite{SM}.

{\it Summary and discussions.}---We have focused on the microscopic
formation of the MO bands in a twisted Hubbard tube capturing the
Q1D nature of K$_2$Cr$_3$As$_3$, a new Q1D multiorbital
superconducting molecular crystal with the moderate Coulomb
interaction and Hund's coupling. A three-channel Tomonaga-Luttinger
liquid Hamiltonian describing the low energy physics of the three
active MO bands ( the $\alpha$, $\beta$, and $\gamma$ bands) is then
derived, showing possible unconventional triplet superconducting
instabilities within a reasonable range of interaction parameters.

The conclusions and implications of our study are compared with the
previous studies \cite{Zhou15,Wu15b} where a phenomenological 3D
Hubbard model for the three active MO bands was proposed based on
the elegant symmetry argument\cite{Zhou15} and investigated by the
random phase approximation\cite{Zhou15,Wu15b} and the mean field
treatment\cite{Wu15b}. First, the twisted structure of the Q1D
[CrAs]$_{\infty}$ tube showing the extended glide reflection
symmetry in accordance with the $C_3$ group is explored in our
approach so that the symmetry property of all the thirty MO bands
(including the three active MO bands) could be identified. Second,
the interactions among the three active MO bands are derived from
the microscopic atomic Hubbard interactions, different from those
proposed phenomenologically. Third, the three diagonal channels in
our Tomonaga-Luttinger Hamiltonian are superpositions of the
original DFT bands, in contrast to the random phase approximation
approach and the mean field treatment.

We found two kinds of spin-triplet pairing instabilities emerging
out from two of the three channels. One involves the Q1D $\alpha$
and $\beta$ bands, another involves all three bands. In the
Luttinger liquid approach, triplet pairing instabilities are due to
gapless spin excitations for $U>J_H$ and $K_c>1$ in the
corresponding channels. The ferromagnetic correlation within the
sublattice of Cr atoms\cite{Jiang14,Wu15b}, though possible, is not
a prerequisite of the triplet states. We also found an intermediate
regime $0.2<J_H/U<0.6$ where the SDW phase emerges. Our solution is
sensitive to the symmetry or regularity of the two conjugated Cr
triangles, seemingly consistent with the recent hydrostatic and
uniaxial pressure experimental study\cite{Wang15}. The exact mapping
from the AOs to MO bands will also pave the way for further
investigations on related effects such as the spin-orbit coupling
within a microscopic framework.

As the present study is limited to the Q1D case, the spatial
symmetry of the superconducting pairing states is not specified.
However, the actual 3D superconductivity can be perceived based on
the Q1D physics because the identification of the three low energy
MO bands is robust owing to the same symmetry argument. The local
interactions among the MO bands are similar to those in the Q1D
case. If the local atomic interactions are estimated as those in
other Cr-based oxides, like SrCrO$_3$\cite{Vaugier12}, one has
$U\sim 2.7\pm 0.5$ eV, $J_H\sim 0.42\pm 0.1$ eV, and $J_H/U\sim
0.16$, then in the Q1D case the channel-2 is in the triplet phase
but close to the SDW low boundary $J_H/U=0.2$ shown in
Fig.\ref{phase2}. The corresponding residual MO interactions
${\tilde U}$ and ${\tilde J}$ in the 3D case are significantly
suppressed, but the ratio ${\tilde J}/{\tilde U}$ enhanced\cite{SM},
corresponding to the regime with small ${\tilde U}$ but relatively
large ${\tilde J}/{\tilde U}$ in Ref.\cite{Zhou15}, where the
triplet $f_{y(3x^2-y^2)}$ pairing state is favored. Of course, we
have not considered the long-range Coulomb interaction and the
electron-phonon coupling, the suppression of residual MO
interactions should necessitate further investigations on these
influences.

Finally, a more intriguing issue is the possible dimensional
crossover from Q1D to 3D which could be tuned by either chemical
substitution\cite{Tang14,Tang15} or physical
pressure\cite{Kong15,Wang15}. On one hand, one of the three active
bands, corresponding to $\nu=1$, evolves with the intertube hopping
and crossovers to the 3D $\gamma$ band which could lead to the line
nodal feature. Meanwhile, the ($\alpha$, $\beta$) bands could remain
in Q1D because the intertube hopping among the AOs with $m=\pm 2$ is
reasonably small. On the other hand, the Q1D superconducting
instability can lead to a true long-range order when the intertube
hopping is taken into account. Recall that the interband triplet
pairing instability in the channel-2 is driven not only by the
$\gamma$ band, but also by the ($\alpha$, $\beta$) bands.
Consequently, the spin-triplet pairing instability in the channel-2
involves both the 3D and Q1D bands. As such a 3D pairing state could
emerge from a normal state of an essentially Q1D Luttinger liquid
characteristic, a scenario which is likely consistent with available
experiments. It is desirable to investigate the related crossover
behavior in this class of materials in the future.

We are grateful to Chao Cao, Guanghan Cao, and Yi Zhou  for
stimulating discussions. This work was supported in part by the NSF
of China (under Grants No. 11274084, No. 11304071, and No.
11474082).

\widetext

\newpage
\clearpage
\renewcommand{\theequation}{S\arabic{equation}}
\setcounter{equation}{0}
\renewcommand{\thetable}{S\arabic{table}}
\setcounter{table}{0}
\renewcommand{\thefigure}{S\arabic{figure}}
\setcounter{figure}{0}

\section*{ "Formation of Molecular-Orbital Bands in a Twisted Hubbard
Tube: Implications for Unconventional Superconductivity in
K$_2$Cr$_3$As$_3$"}

By: Hanting Zhong, Xiao-Yong Feng, Hua Chen, and Jianhui Dai

\section*{SUPPLEMENTAL MATERIAL}

This is the Supplemental Material (SM) for our paper titled
"Formation of Molecular-Orbital Bands in a Twisted Hubbard Tube:
Implications for Unconventional Superconductivity in
K$_2$Cr$_3$As$_3$".\cite{SMZhong15} In this SM, we provide the
detailed solution of the tight-binding Hamiltonian as well as some
supplemental discussions on various related issues, including the
band structure fitting, a comparison with previous theoretical
studies, the one-loop RG equations, and the Luttinger parameters
away from the condition of atomic orbital rotational symmetry.

\subsection{Solution of orbital quasi-degenerate tight-binding Hamiltonian and fitting the band structure}

The tight-binding Hamiltonian discussed in the main text is given by
\begin{eqnarray}
H_0&=&-\sum_{n,\sigma}\sum_{m,m'}\sum_{I\neq I'}\{
t^{(1)}_{mm'}d^{\dagger (m)}_{nI\sigma} d^{(m')}_{nI'\sigma}
+t^{(2)}_{mm'}d^{\dagger(m)}_{n\bar{I}\sigma}
d^{(m')}_{n\bar{I}'\sigma}+t^{(3)}_{mm'}d^{\dagger(m)}_{nI\sigma}
d^{(m')}_{n\bar{I}'\sigma}+t^{(4)}_{mm'}d^{\dagger(m)}_{nI\sigma}
d^{(m')}_{n+1\bar{I}'\sigma} \}\nonumber\\
&&+\sum_{n,\sigma}\sum_{m,m'}\sum_{I}t^{(5)}_{mm'}\{d^{\dagger(m)}_{nI\sigma}
d^{(m')}_{n+1I\sigma}+ d^{\dagger(m)}_{n{\bar I}\sigma}
d^{(m')}_{n+1{\bar
I}\sigma}\}+\sum_{n,\sigma}\sum_{m}\sum_{I}\{E_{m}{n}^{(m)}_{nI\sigma}+
{\bar E}_{m}{n}^{(m)}_{n{\bar I}\sigma}\}.
\end{eqnarray}
Where, $d^{(m)}_{nI\sigma}$ annihilates a Cr $3d$-electron moving
along the $z$-axis at the $n$-unit cell, with spin polarization
$\sigma$ ($=\uparrow,\downarrow$), orbital component $m$ ($=0,\pm
2$), intra-triangle location $I$($=A,B,C$), as well as the conjugate
triangle location $\bar {I}$($=\bar {A},\bar {B},\bar {C}$). Here,
the site locations correspond to $(a,\xi)$ introduced in the main
text as: $A=(1,1)$, $B=(2,1)$, $C=(3,1)$, $\bar {A}=(1,2)$, $\bar
{B}=(2,2)$, $\bar {C}=(3,2)$. The notation $(a,\xi)$ introduced in
the main text is explicit to accommodate the group representation ,
while ($A,B,C$) or ($\bar {A},\bar {B},\bar {C}$) are more
transparent. Either notations will be used for convenience in this
SM. The hopping parameters $t^{(i)}_{mm'}$ ( $i=1,2,3$ and $4$) are
those for the n.n. sites shown in Fig.1. In the present orbital
quasi-degenerate model, these intraorbital and interorbital hopping
parameters satisfy the relationship $t^{(i)}_{mm'}=\eta
t^{(i)}_m\delta_{|m||m'|}$ for $m\neq m'$ with the ratio $|\eta|<1$
( In the following we take $\eta>0$ without losing the generality).
$t^{(5)}_{mm'}=t^{(5)}_{m}\delta_{mm'}$ is the next nearest neighbor
(intraorbital) hopping along the tube direction. The CEF term is
given by $E_{m}$ (or ${\bar E}_{m}$), with ${n}^{(m)}_{nI\sigma}$
(or ${n}^{(m)}_{n{\bar I}\sigma}$) being the corresponding density
operators.

In order to solve the  non-interacting Hamiltonian for a single
cluster [CrAs]$_6$ in the n-th unit cell, $H_{0,n}$,  we introduce
$d^{(m)}_{n}=\left (d^{(m)}_{(n,A)},d^{(m)}_{(n,{\bar
A})},d^{(m)}_{(n,B)},d^{(m)}_{(n,{\bar B})},
d^{(m)}_{(n,C)},d^{(m)}_{(n,{\bar C})} \right)^T$ for each $m=0,\pm
1, \pm 2$ (the spin index $\sigma$ is implied). Because of orbital
mixing between $m=\pm1$ or $\pm 2$, we need to introduce another set
of base ${\tilde d}^{(\pm|m|)}_{n}=\frac{1}{\sqrt 2}[d^{(m)}_{n}\pm
d^{(-m)}_{n}]$. Then all thirty atomic orbitals (MOs) in a unit cell
can be described by ${\tilde d}^{(\tau)}_{n}$, with $\tau=0, \pm 1,
\pm 2$ ( denote ${\tilde d}^{(\tau=0)}_{n}\equiv {d}^{(m=0)}_{n} $)
The MOs of the n-th cluster, ${C}^{(\tau)}_{n}= \left
(c^{(\tau)}_{(n,1,1)},c^{(\tau)}_{(n,
1,2)},c^{(\tau)}_{(n,2,1)},c^{(\tau)}_{(n,2,2)},
c^{(\tau)}_{(n,3,1)},c^{(\tau)}_{(n,3,2)} \right)^T$, are defined as
the base which diagonalizes Hamiltonian $H_{0,n}=\sum_{\tau}
{C}^{\dagger (\tau)}_{n} {{H}^{(\tau)}_{0,n}}{C}^{(\tau)}_{n}$.
Here, $H^{(\tau)}_{0,n}=diag \{{E}^{(\tau)}_{(a,\xi)}\}$ is a
diagonal matrix with eigenvalues ${E}^{(\tau)}_{(a,\xi)}$. The MOs
can be obtained by $C^{(\tau)}_{n}=( {\hat R}\otimes{\hat
Q}_0){\tilde d}^{(\tau)}_{n}$ with $\omega=e^{i\varphi}$,
$\varphi=2\pi/3$, ${\hat R}$ and ${\hat Q}_0$ are given in Eq.(3) in
the main text. Explicitly, we have\cite{SMnote}
\begin{eqnarray}
c^{(\tau)}_{(n,1,1)}&=&\frac{1}{\sqrt 6} \left( {\tilde
d}^{(\tau)}_{(n,A)}+{\tilde d}^{(\tau)}_{(n,{\bar A})}+{\tilde
d}^{(\tau)}_{(n,B)}+{\tilde d}^{(\tau)}_{(n,{\bar B})}+{\tilde
d}^{(\tau)}_{(n,C)}+{\tilde
d}^{(\tau)}_{(n,{\bar C})} \right),\\
c^{(\tau)}_{(n,1,2)}&=&\frac{1}{\sqrt 6} \left( {\tilde
d}^{(\tau)}_{(n,A)}-{\tilde d}^{(\tau)}_{(n,{\bar A})}+{\tilde
d}^{(\tau)}_{(n,B)}-{\tilde d}^{(\tau)}_{(n,{\bar B})}+{\tilde
d}^{(\tau)}_{(n,C)}-{\tilde
d}^{(\tau)}_{(n,{\bar C})} \right),\\
c^{(\tau)}_{(n,2,1)}&=&\frac{1}{\sqrt 6} \left( {\tilde
d}^{(\tau)}_{(n,A)}+{\tilde d}^{(\tau)}_{(n,{\bar A})}+\omega
{\tilde d}^{(\tau)}_{(n,B)}+\omega {\tilde d}^{(\tau)}_{(n,{\bar
B})}+\omega^{-1} {\tilde d}^{(\tau)}_{(n,C)}+\omega^{-1} {\tilde
d}^{(\tau)}_{(n,{\bar C})} \right),\\
c^{(\tau)}_{(n,2,2)}&=&\frac{1}{\sqrt 6} \left( {\tilde
d}^{(\tau)}_{(n,A)}-{\tilde d}^{(\tau)}_{(n,{\bar A})}+\omega
{\tilde d}^{(\tau)}_{(n,B)}-\omega {\tilde d}^{(\tau)}_{(n,{\bar
B})}+\omega^{-1} {\tilde d}^{(\tau)}_{(n,C)}-\omega^{-1} {\tilde
d}^{(\tau)}_{(n,{\bar C})} \right),\\
c^{(\tau)}_{(n,3,1)}&=&\frac{1}{\sqrt 6} \left( {\tilde
d}^{(\tau)}_{(n,A)}+{\tilde d}^{(\tau)}_{(n,{\bar A})}+\omega^{-1}
{\tilde d}^{(\tau)}_{(n,B)}+\omega^{-1} {\tilde
d}^{(\tau)}_{(n,{\bar B})}+\omega {\tilde d}^{(\tau)}_{(n,C)}+\omega
{\tilde
d}^{(\tau)}_{(n,{\bar C})} \right),\\
c^{(\tau)}_{(n,3,2)}&=&\frac{1}{\sqrt 6} \left( {\tilde
d}^{(\tau)}_{(n,A)}-{\tilde d}^{(\tau)}_{(n,{\bar A})}+\omega^{-1}
{\tilde d}^{(\tau)}_{(n,B)}-\omega^{-1} {\tilde
d}^{(\tau)}_{(n,{\bar B})}+\omega {\tilde d}^{(\tau)}_{(n,C)}-\omega
{\tilde d}^{(\tau)}_{(n,{\bar C})} \right).
\end{eqnarray}

In order to solve the whole tight-binding Hamiltonian $H_0$, we need
to introduce $d^{(m)}_{k}=\left (d^{(m)}_{(k,A)},d^{(m)}_{(k,{\bar
A})},d^{(m)}_{(k,B)},d^{(m)}_{(k,{\bar B})},
d^{(m)}_{(k,C)},d^{(m)}_{(k,{\bar C})} \right)^T$, and the
corresponding ${\tilde d}^{(\tau)}_{k}$ in the momentum space. Then,
the Hamiltonian can be diagonalized  by ${C}^{(\tau)}_{k}= \left
(c^{(\tau)}_{(k,1,1)},c^{(\tau)}_{(k,
1,2)},c^{(\tau)}_{(k,2,1)},c^{(\tau)}_{(k,2,2)},
c^{(\tau)}_{(k,3,1)},c^{(\tau)}_{(k,3,2)} \right)^T$, in the form of
$H_{0}=\sum_{\tau,k} {C}^{\dagger (\tau)}_{k}
{{H}^{(\tau)}_{0,k}}{C}^{(\tau)}_{k}$. Here, $H^{(\tau)}_{0,k}=diag
\{{\cal E}^{(\tau)}_{(a,\xi)}(k)\}$ is the diagonal matrix with
eigenvalues ${\cal E}^{(\tau)}_{(a,\xi)}(k)$ given by
\begin{eqnarray}
{\cal E}^{(\tau)}_{(a,\xi) }(k)=-\frac{\lambda^{(\tau)}_{a}}{2}
\left[\epsilon^{(\tau)}_{(a,1)}+\epsilon^{(\tau)}_{(a,2)}
+(-1)^{\xi}\sqrt{(\epsilon^{(\tau)}_{(a,1)}-\epsilon^{(\tau)}_{(a,2)})^2
+4\rho_{\tau}^2}\right].
\end{eqnarray}
In above, $k$ is the crystal momentum along the tube direction,
$\lambda^{(\tau)}_a=[1+sign(\tau)|\eta|]\lambda_a$ for $a=1,2,3$,
with $\lambda_1=2$ and $\lambda_2=\lambda_3=-1$ the eigenvalues of
$\hat R$,
$\epsilon^{(\tau)}_{(a,\xi)}=t^{(\xi)}_\tau+\frac{2t^{(5)}_\tau\cos
k-E^{(\xi)}_\tau-\mu}{\lambda^{(\tau)}_a}$,
$\rho_{\tau}=|t^{(3)}_\tau+t^{(4)}_\tau e^{ik}|$,
and $\mu$ the chemical potential.

Accordingly, the electron operators in MO bands are given by
$C^{(\tau)}_{k} = {\hat R}\otimes diag\{{\hat Q}^{(\tau)}_a(k) \}
{\tilde d}^{(\tau)}_{k}$. Here, $diag\{{\hat Q}^{(\tau)}_a(k) \}$ is
a direct product of sub-matrices ${\hat Q}^{(\tau)}_a(k)$ defined
for each eigenstates $\lambda_a$ of the $C_3$ rotation as given by
\begin{eqnarray}
{\hat Q}^{(\tau)}_{a}(k)= \left(
\begin{array}{cc}
 \cos\alpha^{(\tau)}_{(a,1)}&\sin\alpha^{(\tau)}_{(a,1)}e^{-i\theta_{\tau}} \\
 \sin\alpha^{(\tau)}_{(a,2)}e^{i\theta_{\tau}}&\cos\alpha^{(\tau)}_{(a,2)}
\end{array}
\right),
\end{eqnarray}
with $\tan \theta_{a\tau}=\frac{t^{(4)}_\tau\sin
k}{t^{(3)}_\tau+t^{(4)}_\tau\cos k}$, $\cos\alpha^{(\tau)}_{(a,\xi)
}=\frac{1}{\sqrt{ 1+(\Delta^{(\tau)}_{(a,\xi)})^2}}$, and
$\Delta^{(\tau)}_{(a,\xi)}=\frac{{\cal
E}^{(\tau)}_{(a,\xi)}-\epsilon^{(\tau)}_{(a,\xi)}}{\rho_{a\tau}}$.

The Fourier transformation of ${C}^{(\tau)}_{k}$ back to the spatial
space does not return exactly, though similar, to the forms as
defined by Eq.(S2-S7), because the corresponding coefficients in
each terms are now $k$-dependent due to the intercell coupling along
the tube direction. This feature will in general lead to various
long-range electron correlations among the MOs of different unit
cells. However, upon summation over the whole all unit cells, the
slowly varying terms dominate the contributions. So as long as only
the local interactions of MO bands are concerned, ${\hat Q}_{0}$ or
Eqs.(S2-S7) can be used to deduce these interactions as given in the
next section.

In order to fit the DFT band structure, we re-express the
eigenvalues in a more explicit form
\begin{eqnarray}
{\cal E}^{(\tau)}_{(a,\xi)}(k)&=&-2t^{(5)}_\tau\cos k+\mu
+\frac{E_\tau+{\bar E}_\tau}{2}\nonumber\\
&&-\frac{[1+sign(\tau)\eta]\lambda_a}{2}\left\{t^{(1)}_\tau+
t^{(2)}_\tau +(-1)^{\xi} \sqrt{(t^{(1)}_\tau-
t^{(2)}_\tau-\frac{E_\tau-{\bar
E}_\tau}{[1+sign(\tau)\eta]\lambda_a})^2
+4|t^{(3)}_\tau+t^{(4)}_\tau e^{ik}|^2}\right\}.
\end{eqnarray}
As explained in the main text, we fit the three active DFT bands
$\alpha$, $\beta$, and $\gamma$ along the $\Gamma$-A direction ( the
tube direction or the c-axis)\cite{SMJiang14,SMWu15}. The
non-degenerated $\gamma$ band is contributed mainly from the
$d_{z^2}$ orbital with $m=0$, is fitted by the singlet MO band
indexed by $(\tau=0,a=1,\xi=1)$, with the energy ${\cal
E}^{(0)}_{1,1}(k)$,
\begin{eqnarray}
{\cal E}_{(\gamma)}(k)=-2t^{(5)}_0\cos k+\mu+\frac{E_0+{\bar
E}_0}{2}-t^{(1)}_0- t^{(2)}_0 +\sqrt{(t^{(1)}_0-
t^{(2)}_0-\frac{E_0-{\bar E}_0}{2})^2 +4|t^{(3)}_0+t^{(4)}_0
e^{ik}|^2}.
\end{eqnarray}
The $\alpha$- and $\beta$-bands, which are degenerate along the
$\Gamma$-A direction, are fitted by  the MO bands indexed by $(
\tau=-2, a=2,\xi=2)$ and $(\tau=-2,a=3,\xi=2)$, with the energy
${\cal E}^{(-2)}_{(2,2)}(k)={\cal E}^{(-2)}_{(3,2)}(k)$
\begin{eqnarray} {\cal
E}_{(\alpha,\beta)}(k)=-2t^{(5)}_{2}\cos k+\mu+\frac{E_{2}+{\bar
E}_{2}}{2}+(1+\eta)\frac{t^{(1)}_{2}+ t^{(2)}_{2}}{2}
+\frac{1+\eta}{2}\sqrt{(t^{(1)}_{2}- t^{(2)}_{2}+\frac{E_{2}-{\bar
E}_{2}}{1+\eta})^2 +4|t^{(3)}_{2}+t^{(4)}_{2} e^{ik}|^2}.
\end{eqnarray}
Both Eqs.(S11,S12) take the form as $ {\cal E}(k) \propto
A_{\tau}+B_{\tau}\cos k+\sqrt{C_{\tau}+D_{\tau}\cos^2\frac{k}{2} }
$, each with four independent coefficients $A_{\tau}$, $B_{\tau}$,
$C_{\tau}$, and $D_{\tau}$. For the $\gamma$-band,
$A_0=\mu+\frac{E_0+{\bar E}_0}{2}-t^{(1)}_0- t^{(2)}_0 $,
$B_0=-2t^{(5)}_0$, $C_0=(t^{(1)}_0-t^{(2)}_0-\frac{E_0-{\bar
E}_0}{2})^2+4(t^{(3)}_0-t^{(4)}_0)^2$, $D_0=16t^{(3)}_0t^{(4)}_0$.
The best fitting is given by $A_0=-1.306$ eV, $B_0=1.049$ eV,
$C_0=2.194$ eV, and $D_0=-2.121$ eV. For the $\alpha$- and
$\beta$-bands, $A_2=\mu+\frac{E_2+{\bar
E}_2}{2}+\frac{1+\eta}{2}(t^{(1)}_2+t^{(2)}_2)$, $B_2=-2t^{(5)}_2$,
$C_2=\frac{(1+\eta)^2}{4}[(t^{(1)}_2-t^{(2)}_2-\frac{E_2-{\bar
E}_2}{2})^2+4(t^{(3)}_2-t^{(4)}_2)^2]$,
$D_2=4(1+\eta)^2t^{(3)}_2t^{(4)}_2$. The best fitting is given by
$A_2=-1.500$ eV, $B_2=0.962$ eV, $C_2=3.450$ eV, and $D_2=-3.074$
eV. The negative $D_{\tau}$ implies opposite signs of
$t^{(3)}_{\tau}$ and $t^{(4)}_{\tau}$. The relatively large value of
$C_{\tau}$ implies a sizable difference in length or electron
occupation between the two conjugated triangles as already indicated
in the DFT calculations.\cite{SMJiang14,SMWu15}

Of course, the above fitting is by no means rigorous, given the fact
that the renormalization effect may be not adequately accounted in
the DFT band structure. It is also possible to fit the DFT band
structure within a reasonable approximation by other sets of
parameters. For instance, we can use relatively smaller parameters
$B_{(\tau)}$ and $C_{(\tau)}$, but positive $D_{\tau}$, the overall
lineshape and band width are still closed to the DFT results. In
comparison with the bare tight-binding parameters these fitting
parameters should be all effective after renormalization. As far as
the three active MO bands $\alpha,\beta$ and $\gamma$ are concerned,
the fitting formulae involve eight independent coefficients. Because
the total number of free tight-binding parameters used in fitting
exceeds eight, we cannot determine these parameters uniquely. On the
other hand, the precise values of these parameters are not important
in our present study. As we have shown in the main text, only the
symmetry property of the active MOs and local interactions between
them play the most crucial role in the resultant Luttinger liquid
theory.

\subsection{Molecular orbital interaction parameters and comparison with previous theoretical studies}

In the main text of this paper, we have considered the microscopic
atomic orbital(AO) Hubbard model with the local intraorbital Coulomb
interaction $U$ and interorbital Coulomb interaction $U'$, Hund's
coupling $J_H$, and pair hopping $J_p$, among all five atomic
3d-orbitals $m=0,\pm 1, \pm 2$. Explicitly, the interaction matrices
take the following forms
\begin{eqnarray}
\left( \begin{array}{ccccc}
 U &  U' & U' & U'& U'\\
 U'&  U  & U' & U'& U'\\
 U'&  U' & U  & U'& U'\\
 U'&  U' & U' & U & U'\\
 U'&  U' & U' & U'& U \\
  \end{array}
\right), ~~~\left(
\begin{array}{ccccc}
 0 & J_H & J_H & J_H & J_H \\
 J_H & 0 & J_H & J_H & J_H \\
 J_H & J_H & 0 & J_H & J_H \\
J_H & J_H & J_H & 0 & J_H \\
J_H & J_H & J_H & J_H & 0
\end{array} \right),~~~\left(
\begin{array}{ccccc}
 0 & J_p & J_p & J_p & J_p \\
 J_p & 0 & J_p & J_p & J_p \\
 J_p & J_p & 0 & J_p & J_p \\
 J_p & J_p & J_p & 0 & J_p \\
 J_p & J_p & J_p & J_p & 0
\end{array}
\right).
\end{eqnarray}

Using the inverse mapping from the MOs to AOs, various two-particle
interactions among the MOs are induced by the above local AO
interactions. In general, the induced MO interactions are non-local
due to the inter-cell hopping. As far as the local MO interactions
are focused, the matrix ${\hat Q}_0$ or Eqs.(S2-S7) can be used in
deducing the local MO interactions given in Eq.(5) in the main text,
as the short-range interactions are mainly due to the slowly-varying
part. Such the local MO interactions depend on local AO interactions
via various products taking the forms like

\begin{eqnarray}
d_{(n,a,\xi),\sigma_1}^{\dag(\tau_1)}d_{(n,a,\xi),\sigma_2}^{\dag(\tau_2)}d_{(n,a,\xi),\sigma_3}^{(\tau_3)}d_{(n,a,\xi),\sigma_4}^{(\tau_4)}
~~~~~~~~~~~~~~~~~~~~~~~~~~~~~~~~~~~~~~~~~~~~~~~~~~~~~~~~
~~~~~~~~~~~~~~~~~~~~~~~~~~~~~~~~~\\
=\sum_{{a_i},{\xi_i}}[{\hat Q}_0^{-1}\otimes{\hat
R}^{-1}]^{\dagger}_{(\tau_1,a_1,\xi_1)} \cdot[{\hat
Q}_0^{-1}\otimes{\hat R}^{-1}]^{\dagger}_{(\tau_2,a_2,\xi_2)}
\cdot[{\hat Q}_0^{-1}\otimes{\hat
R}^{-1}]_{(\tau_3,a_3,\xi_3)}\cdot[{\hat Q}_0^{-1}\otimes{\hat
R}^{-1}]_{(\tau_4,a_4,\xi_4)}\nonumber\\
C_{(n,a_1,\xi_1),\sigma_1}^{\dag(\tau_1)}C_{(n,a_2,\xi_2),\sigma_2}^{\dag(\tau_2)}C_{(n,a_3,\xi_3),\sigma_3}^{(\tau_3)}C_{(n,a_4,\xi_4),\sigma_4}^{(\tau_4)}
.~~~~~~~~~~~~~~~~~ ~~~~~~~~~~~~~~~~~~~~~~~~~~~~~~~~~\nonumber
\end{eqnarray}

In the previous section, by solving the tight-binding Hamiltonian
and fitting the DFT band structure, the three active molecular
orbital bands $\nu=1,2,3$ are identified. Based on the symmetry
argument, $\nu=1$ corresponds to the $\gamma$-band with $m=0$,
denoted by the molecular orbital electron annihilation operator $
c_{1}(n)\equiv c^{(0)}_{(n,1,1)}$ (with $\tau=0, a=1,\xi=1$) in
Eq.(S2). While $\nu=2$ and $\nu=3$ correspond to the $\alpha$- and
$\beta$-bands with $m=\pm 2$ , denoted by the molecular orbital
electron annihilation operator $c_{2}\equiv c^{(-2)}_{(n,2,2)}$
(with $\tau=-2, a=2,\xi=2$) and $c_{3}\equiv c^{(-2)}_{(n,3,2)}$
(with $\tau=-2, a=3,\xi=2$) in Eqs.(S5) and (S7), respectively.
After some tedious but straightforward algebras thanks to symmetry
properties of the matrices ${\hat R}$ and ${\hat Q}_0$, the
following interaction matrices for various two-particle Coulomb
interactions and Hund's couplings for the above three active
molecular orbitals are deduced:
\begin{eqnarray}
{\tilde U}&=& \left(
\begin{array}{ccc}
 {\tilde U}_1 &  {\tilde U}_{12} & {\tilde U}_{13} \\
 {\tilde U}_{21} & {\tilde U}_{2} & {\tilde U}_{23} \\
 {\tilde U}_{31} & {\tilde U}_{32} & {\tilde U}_{3}
\end{array}
\right) = \left(
\begin{array}{ccc}
 \frac{U}{6} & \frac{U'}{12}-\frac{J_H}{24} &  \frac{U'}{12}-\frac{J_H}{24} \\
  \frac{U'}{12}-\frac{J_H}{24} & \frac{U+U'+J_H+J_p}{12} & \frac{U+U'+J_H+J_p}{48} \\
 \frac{U'}{12}-\frac{J_H}{24} & \frac{U+U'+J_H+J_p}{48} & \frac{U+U'+J_H+J_p}{12}
\end{array}
\right), \\
{\tilde J}&=& \left(
\begin{array}{ccc}
 0 &  {\tilde J}_{12} & {\tilde J}_{13} \\
 {\tilde J}_{21} & 0 & {\tilde J}_{23} \\
 {\tilde J}_{31} & {\tilde J}_{32} & 0
\end{array}
\right) = \left(
\begin{array}{ccc}
 0 & \frac{J_H}{6} &  \frac{J_H}{6} \\
  \frac{J_H}{6} & 0 & \frac{U+U'+J_H+J_p}{12} \\
 \frac{J_H}{6} & \frac{U+U'+J_H+J_p}{12} & 0
\end{array}
\right).
\end{eqnarray}

In addition, a three-band interaction term
\begin{eqnarray}
{\tilde J}_{123}\left[
c^{\dagger}_{1\uparrow}(n)c^{\dagger}_{1\downarrow}(n)
c_{2\downarrow}(n)c_{3\uparrow}(n)+(2\leftrightarrow 3)+h.c.\right]
\end{eqnarray}
emerges, with ${\tilde J}_{123}=J_p/6$. Note that this peculiar
interaction is absent in the previous theoretical
studies\cite{SMZhou15,SMWu15b}. However, as we shall show in the
next section, this term does not influence the instabilities we
concern.

Usually, we assume the orbital rotational symmetry of the
interacting Hamiltonian in the atomic orbital Hubbard model, i.e.,
$U'=U-2J_H$ and $J_p=J_H$.\cite{SMCastellani78} For a given
material, it is understood that values of these local interactions
can be meaningfully determined when band structure calculations as
using the constraint density functional approaches are implemented
by a complete description of the corresponding atomic Hubbard model.
So far such detailed calculations on the compound
K$_2$Cr$_{3}$As$_{3}$ are not yet available. However, these local
atomic interactions can be inferred from those in other Cr-based
oxides, like SrCrO$_3$. According to Ref. \cite{SMVaugier12},  one
has $U\sim 2.7$ eV, $J_H\sim 0.42$ eV in SrCrO$_3$. These values are
relatively smaller than but still closed to those of Fe$^{2+}$
systems like SrFeO$_2$ or iron pnictides, where $U\sim 3-5$ eV,
$J_H\sim 0.50-0.70$ eV. \cite{SMXiang08} The ratio $J_H/U$ of these
systems does not change too much. Hence as a rough estimate, we
assume the similar values of SrCrO$_3$ for the present compound,
obtaining the local molecular orbital interaction matrices from
Eqs.(S15,S16):
\begin{eqnarray}
{\tilde U} \approx   \left(
\begin{array}{ccc}
 0.45 & 0.14 & 0.14 \\
 0.14 & 0.45 & 0.11 \\
 0.14 & 0.11 & 0.45
\end{array}
\right), ~~~{\tilde J} \approx  \left(
\begin{array}{ccc}
 0 & 0.07 & 0.07 \\
 0.07 & 0 & 0.45 \\
 0.07 & 0.45 & 0
\end{array}
\right).
\end{eqnarray}

Notice that it is always possible for these values to be fluctuated
within 20\% or even more for a given material by using different
approaches. With this understanding, it is interesting to compare
our results with those in Refs.\cite{SMZhou15,SMWu15b}.  In our Q1D
Luttinger theory, while the channel "3" is always critical for a
much wider regime $U>J_H$, the channel "2" is expected to be located
at the spin triplet phase because $J_H/U\sim 0.16$. However, this
triplet state should close to the border of the SDW phase,
$J_H/U=0.2$, as shown in Fig.3 in the main text.

In Ref.\cite{SMZhou15}, the following interaction matrices for
phenomenological molecular bands are introduced:
\begin{eqnarray}
{\tilde U}_{\rm Z}= \left(
\begin{array}{ccc}
 U_1 &  U_2 & U'_2 \\
 U_2 &  U_1 & U'_2 \\
 U'_2 & U'_2& U'_1
\end{array}
\right), ~~~{\tilde J}_{\rm Z}= \left(
\begin{array}{ccc}
 0 & J & J' \\
 J & 0 & J' \\
 J' & J' & 0
\end{array}
\right), ~~~{\tilde J}_p={\tilde J}_{\rm Z},
\end{eqnarray}
while in Ref.\cite{SMWu15b},
\begin{eqnarray}
{\tilde U}_{\rm W}=\left(
\begin{array}{ccc}
 U &  V & V \\
 V &  U & V \\
 V &  V & U
\end{array}
\right), ~~~{\tilde J}_{\rm W}= \left(
\begin{array}{ccc}
 0 & J_H & J_H \\
 J_H & 0 & J_H\\
 J_H & J_H & 0
\end{array}
\right), ~~~{\tilde J}_p={\tilde J}_{\rm W}.
\end{eqnarray}
Here, we use the same notations introduced in
Refs.\cite{SMZhou15,SMWu15b}, respectively. The relationships
$U'_1=U_1$, $U'_2=U_2$, and $0.5<J'/J<2.0$ were used in calcuations
\cite{SMZhou15}. It should be understood that (i) the parameters
used in these studies are for the molecular orbital bands; (ii) they
do not rigorously correspond to our interaction matrices given by
Eqs.(S15-S17). Nevertheless, we may expect that if the largest
elements in the interaction matrices ${\tilde U}$ and ${\tilde J}$
dominate in the random phase approximation (RPA) approach, the
values of atomic interactions estimated in our case correspond to
the case of $U_1=0.45$, $J=0.45$, and $J/U_1=1$ in
Ref.\cite{SMZhou15}. So the channel "2" should correspond to the
regime with small $U$ but large $J/U$ in Ref.\cite{SMZhou15}. As
shown in Fig. 3(a) in Ref.\cite{SMZhou15}, this phase favors the
spin triplet $f_{y(3x^2-y^2)}$ pairing state. Notice that these
values are out of the calculated regime of the phase diagram Fig (6)
in Ref.\cite{SMWu15b}, but the same tendency is reached if this
phase diagram is extended to large ration of $J/U$.

\subsection{One-loop RG equations and absence of the three-band instability}

Owing to the symmetry between the MO bands $\nu=2$ and $\nu=3$,
there are twelve permissible perturbations which can be classified
in terms of the formal $g$-ology description
\cite{SMSolyom79,SMGiamarchi04} using the same notation in
\cite{SMKrotov97,SMGonzalez05,SMCarpentier06}:

\begin{eqnarray}\label{parameters}
{\cal H}_{\rm int} &=& g^{(1)}\sum_{\sigma\sigma'}R^{\dagger}_{1\sigma}L_{1\sigma}L^{\dagger}_{1\sigma'}R_{1\sigma'}\nonumber\\
&+&g^{(2)}\sum_{\sigma\sigma'}R^{\dagger}_{1\sigma}R_{1\sigma}L^{\dagger}_{1\sigma'}L_{1\sigma'}\nonumber\\
&-&g^{(1)}_{1}\sum_{\sigma\sigma'}R^{\dagger}_{2\sigma}R_{2\sigma'}L^{\dagger}_{3\sigma'}L_{3\sigma}+(2\leftrightarrow 3)\nonumber\\
&-&g^{(2)}_{1}\sum_{\sigma\sigma'}R^{\dagger}_{2\sigma}L_{2\sigma'}L^{\dagger}_{3\sigma'}R_{3\sigma}+(2\leftrightarrow 3)\nonumber\\
&+&g^{(1)}_{2}\sum_{\sigma\sigma'}R^{\dagger}_{2\sigma}L_{2\sigma}L^{\dagger}_{3\sigma'}R_{3\sigma'}+(2\leftrightarrow 3)\nonumber\\
&+&g^{(2)}_{2}\sum_{\sigma\sigma'}R^{\dagger}_{2\sigma}R_{2\sigma}L^{\dagger}_{3\sigma'}L_{3\sigma'}+(2\leftrightarrow 3)\nonumber\\
&+&g^{(1)}_{4}\sum_{\sigma\sigma'}R^{\dagger}_{2\sigma}L_{2\sigma}L^{\dagger}_{2\sigma'}R_{2\sigma'}+(2\leftrightarrow 3)\nonumber\\
&+&g^{(2)}_{4}\sum_{\sigma\sigma'}R^{\dagger}_{2\sigma}R_{2\sigma}L^{\dagger}_{2\sigma'}L_{2\sigma'}+(2\leftrightarrow 3)\nonumber\\
&-&f^{(1)}\sum_{\sigma\sigma'}[R^{\dagger}_{1\sigma}R_{1\sigma'}L^{\dagger}_{2\sigma'}L_{2\sigma}+(2\leftrightarrow 3)]+(R\leftrightarrow L)\nonumber\\
&+&f^{(2)}\sum_{\sigma\sigma'}[R^{\dagger}_{1\sigma}R_{1\sigma}L^{\dagger}_{2\sigma'}L_{2\sigma'}+(2\leftrightarrow 3)]+(R\leftrightarrow L)\nonumber\\
&+&u\sum_{\sigma\sigma'}[R^{\dagger}_{1\sigma}L^{\dagger}_{1\sigma'}L_{2\sigma'}R_{3\sigma}+(2\leftrightarrow 3)]+h.c.\nonumber\\
&+&v\sum_{\sigma\sigma'}[R^{\dagger}_{1\sigma}L^{\dagger}_{1\sigma'}R_{2\sigma'}L_{3\sigma}+(2\leftrightarrow
3)]+h.c..
\end{eqnarray}

\begin{figure}
\includegraphics[width=12.0cm]{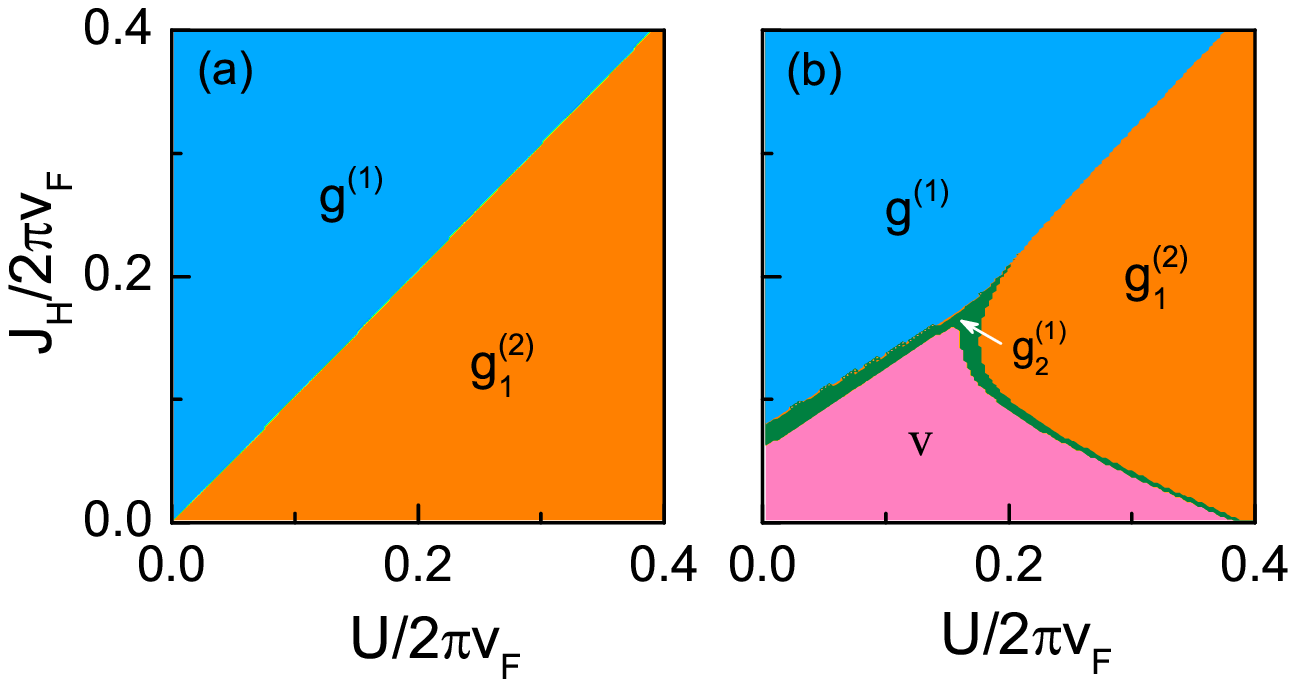}
\caption{The dominating perturbations in the one-loop RG equations.
(a)All the initial parameters are given by Eqs.(S15-S17,S22); (b)
The initial values of $u$ and $v$ have the opposite sign using
$v/2\pi v_F = J_H/6-0.02$ and $u/2\pi v_F=J_H/6$ for small $J_H$,
while the other parameters are still given by Eqs. (S15-S17,S22). }
\label{RG}
\end{figure}

In this notation, the degeneracy between the MO bands $\nu=2$ and
$\nu=3$ are explicit. The one-loop renormalization group (RG)
equations for these coupling constants evolving with increasing
scaling parameter $l$ are of common type, given by
\cite{SMKrotov97,SMGonzalez05,SMCarpentier06}:
\begin{eqnarray*}
\partial_l\tilde{g}^{(1)}&=&-2(\tilde{g}^{(1)})^2-4\tilde{u}\tilde{v}\\
\partial_l\tilde{g}^{(2)}&=&-(\tilde{g}^{(1)})^2-2\tilde{u}^2-2\tilde{v}^2\\
\partial_l\tilde{g}^{(1)}_1&=&-2(\tilde{g}^{(1)}_1)^2-\tilde{g}^{(2)}_1\tilde{g}^{(1)}_2-2\tilde{u}\tilde{v}\\
\partial_l\tilde{g}^{(2)}_1&=&-2\tilde{g}^{(1)}_1\tilde{g}^{(1)}_2-\tilde{g}^{(2)}_1\tilde{g}^{(2)}_2+\tilde{g}^{(2)}_1\tilde{g}^{(2)}_4-\tilde{u}^2-\tilde{v}^2\\
\partial_l\tilde{g}^{(1)}_2&=&-2\tilde{g}^{(1)}_1\tilde{g}^{(2)}_1-2\tilde{g}^{(1)}_2\tilde{g}^{(2)}_2+\tilde{g}^{(2)}_1\tilde{g}^{(1)}_4-4\tilde{g}^{(1)}_2\tilde{g}^{(1)}_4+2\tilde{g}^{(1)}_2\tilde{g}^{(2)}_4-2\tilde{u}\tilde{v}\\
\partial_l\tilde{g}^{(2)}_2&=&-(\tilde{g}^{(1)}_1)^2-(\tilde{g}^{(2)}_1)^2-(\tilde{g}^{(1)}_2)^2-\tilde{u}^2-\tilde{v}^2\\
\partial_l\tilde{g}^{(1)}_4&=&2\tilde{g}^{(2)}_1\tilde{g}^{(1)}_2-2(\tilde{g}^{(1)}_2)^2-2(\tilde{g}^{(1)}_4)^2\\
\partial_l\tilde{g}^{(2)}_4&=&(\tilde{g}^{(2)}_1)^2-(\tilde{g}^{(1)}_4)^2\\
\partial_l\tilde{f}^{(1)}&=&-2(\tilde{f}^{(1)})^2+2\tilde{u}\tilde{v}-2\tilde{v}^2\\
\partial_l\tilde{f}^{(2)}&=&-(\tilde{f}^{(1)})^2+\tilde{u}^2\\
\partial_l\tilde{u}&=&(2\tilde{f}^{(2)}-\tilde{g}^{(2)}_1-\tilde{g}^{(2)}-\tilde{g}^{(2)}_2)\tilde{u}-(\tilde{f}^{(1)}+\tilde{g}^{(1)}_1+\tilde{g}^{(1)}_2)\tilde{v}\\
\partial_l\tilde{v}&=&-(-2\tilde{f}^{(1)}+\tilde{g}^{(1)}+\tilde{g}^{(1)}_1+\tilde{g}^{(1)}_2)\tilde{u}-(4\tilde{f}^{(1)}-2\tilde{f}^{(2)}+\tilde{g}^{(2)}_1+\tilde{g}^{(2)}+\tilde{g}^{(2)}_2)\tilde{v}.
\end{eqnarray*}
In above, $\tilde{g}\equiv\frac{g}{2\pi v_F}$, and
\begin{eqnarray}
\begin{array}{ll}
  g^{(1)}=\tilde{U}_1, & g^{(2)}=\tilde{U}_1, \\
  g^{(1)}_1=\tilde{J}_{23}, & g^{(2)}_1=\tilde{J}_{23}, \\
  g^{(1)}_2=2\tilde{U}_{23}+\frac{1}{2}\tilde{J}_{23}, & g^{(2)}_2=2\tilde{U}_{23}+\frac{1}{2}\tilde{J}_{23}, \\
  g^{(1)}_4=\tilde{U}_{2}, & g^{(2)}_4=\tilde{U}_{2}, \\
  f^{(1)}=\tilde{J}_{12}, & f^{(2)}=2\tilde{U}_{12}+\frac{1}{2}\tilde{J}_{12},\\
  u=\tilde{J}_{123}, & v=\tilde{J}_{123}.
\end{array}
\end{eqnarray}
The corresponding bare MO interaction parameters, which are related
to initial values of perturbations defined in Eq.(S22) in solving
the RG equations, are given in Eqs.(S15-S17). The initial values of
perturbations $u$ and $v$ are the same in the present model, both
induced by the three-band interaction $\tilde{J}_{123}$. When $l$
approaches a sufficient large cutoff scale $l^* $, the generic
asymptotic solutions take the form $g^{(j)}_i\propto
(l^*-l)^{-\lambda_{j,i}}$. An instability takes place whenever some
of the couplings diverge at a finite scale length $l^*$. We
numerically determine the most divergent coupling when $l$
approaches $l^*$, which is taken to be unit, from below. For usual
$3d$-transition metals, $0<J_H/U<1$, we find that $g^{(1)}$ or
$g^{(2)}_1$ dominates in the large or small $J_H/U$ regimes,
respectively, as plotted in Figure \ref{RG}(a). These two regimes
correspond to the known single-band and the two-band instabilities
respectively\cite{SMKrotov97,SMGonzalez05,SMCarpentier06}.

It was suggested that in addition to these one-band and two-band
instabilities, there may be a new instability driven by the presence
of all three bands due to the perturbation $u$ or $v$. We have
checked that this instability does not occur if the initial $u$ and
$v$ have the same sign as determined in the present case. We have
also checked that if the initial values of $u$ and $v$ have the
opposite sign, say, assuming $v/2\pi v_F = J_H/6-0.02$ and $u/2\pi
v_F=J_H/6$ for very small $J_H$, there is a regime where $v$
dominates over all other perturbations as shown in Fig. \ref{RG}(b).
It is interesting to recall that the electron-phonon coupling, which
has not been adequately considered in the present study, may result
in deviations of effective perturbations away from the initial
values determined in Eq.(S22). How such modification upon
electron-phonon coupling takes place and whether it influences the
superconductivity in the K$_2$Cr$_3$As$_3$ compound deserve further
investigations.

\subsection{The Luttinger parameters away from the condition of atomic orbital rotational symmetry}

The rotational symmetry in the interacting part of a given system is
frequently assumed in literatures, by using the relationships
$J_p=J_H$ and $U'=U-J_H-J_p$ at the level of atomic
orbitals\cite{SMCastellani78}. The relationships are adopted in
plotting the phase diagram Fig.3 in the main text. It is possible
that in realistic and complicated systems this symmetry could be
broken, or the above relationships could not be respected. In order
to understand whether our results are still valid in this case, we
consider a small deviation $\Delta U$ away from the rotational
symmetry, by assuming $U'=U+\Delta U-2J_H$. It is straightforward to
show that the diagonal channels are similar to the case  with
$\Delta U =0$ as given by Eq.(7) in the main text, while the model
parameters appear in the Tomonaga-Luttinger Hamiltonian Eq.(8) are
given by
\begin{eqnarray*}
  t_{c1}=\frac{v_F}{2}+\frac{\tilde{U}_1 }{2\pi},~~ t_{c2}=\frac{v_F}{2}+\frac{\tilde{U}_2 }{2\pi}, ~~a_c= \frac{2\tilde{U}_{12}}{\pi}, ~~b_c =  \frac{2\tilde{U}_{23}}{\pi},\\
  t_{s1}=\frac{v_F}{2}-\frac{\tilde{U}_1 }{2\pi},~~t_{s2}=\frac{v_F}{2}-\frac{\tilde{U}_2 }{2\pi}, ~~a_s= -\frac{\tilde{J}_{12}}{2\pi}, ~~b_s = -\frac{\tilde{J}_{23}}{2\pi}.
\end{eqnarray*}
Then, the Luttinger parameters are given by
$K_{\gamma,i}=\sqrt{\frac{v_F}{2\lambda_{\gamma,i}}}$, with
\begin{eqnarray*}
\lambda_{\gamma,1}&=&\frac{1}{2}(t_{\gamma1}+t_{\gamma2})+\frac{1}{2}\left(b_{\gamma}+\sqrt{8a_{\gamma}^2+(b_{\gamma}+t_{\gamma2}-t_{\gamma1})^2}\right),\nonumber\\
\lambda_{\gamma,2}&=&\frac{1}{2}(t_{\gamma1}+t_{\gamma2})+\frac{1}{2}\left(b_{\gamma}-\sqrt{8a_{\gamma}^2+(b_{\gamma}+t_{\gamma2}-t_{\gamma1})^2}\right),\\
\lambda_{\gamma,3}&=&t_{\gamma2}-b_{\gamma}.
\end{eqnarray*}
Explicitly, we have three pairs of charge and spin Luttinger
parameters in the corresponding channels:
\begin{eqnarray}
&&K_{c,1}=\frac{1}{\sqrt{1+G_1+G_2}},~~ K_{c,2}=\frac{1}{\sqrt{1+G_1-G_2}}, ~~K_{c,3}=1, \nonumber\\
&& K_{s,1}=\frac{1}{\sqrt{1-G_1+G_3}},~~
K_{s,2}=\frac{1}{\sqrt{1-G_1-G_3}},~~ K_{s,3}=1.
\end{eqnarray}
Where,
\begin{eqnarray}
G_1&=&\frac{U}{4\pi v_F}+\frac{\Delta U}{12\pi
v_F},\nonumber\\
G_2&=&\sqrt{8\left(\frac{U+\Delta U}{6\pi v_F}-\frac{5J_H}{12\pi v_F}\right)^2+\left(\frac{U+\Delta U}{12\pi v_F}\right)^2},\nonumber\\
G_3&=&\sqrt{8\left(\frac{J_H}{12\pi
v_F}\right)^2+\left(\frac{U}{12\pi v_F}\right)^2}.
\end{eqnarray}
The fact that the channel-"3" is still in the critical phase
 with $K_{s,3}=K_{c,3}=1$ is apparently due to the degeneracy of the
 MO bands $\nu=2$ and $\nu=3$. Based on these expressions, we find that all our results in the main
text remain unchanged because the role of $\Delta U$ is to modify
the value of U in a simple manner.


\begin{thebibliography}{99}
\bibitem{note1}
The glide reflection symmetry is exact when the two conjugated
triangles are equivalent and the corresponding hopping parameters
and the CEF splittings in the Hamiltonian Eq.(1) are the same.
\bibitem{note2}
The assumption of quasidegenerate AOs leads to a fixed AO content of
each MO band along the $\Gamma\rightarrow A$ direction up to a
rotation around the $z$ axis. The variation of the AO contents does
not alter the symmetry property and local interactions of the MOs.
\bibitem{note3}
In the Luttinger liquid approach the electronic interactions among
various MO bands impact the kinetic Hamiltonian after bosonization,
leading to reconstruction of the original DFT bands. So we shall use
the concept "channel" for the three diagonal quasiparticle bands as
defined in Eq.(7).
\bibitem{note4}
The three-band instability discussed in Ref.\cite{Carpentier06} may
appear when the electron-phonon coupling is relatively larger than
the pair hopping or the Hund's coupling.
\bibitem{note5}
The present study does not exclude the possibility of a spin singlet
pairing due to the critical nature of the channel 3.
\end{thebibliography}
\begin{thebibliography}{99}
\bibitem{SMnote} There could be a prefactor $(-1)$ for the sectors
$\lambda_a=-1$ (for $a=2,3$) related to the sector $\lambda_a=2$
(for $a=1$) as defined in Eq.(S9). This global prefactor can be
absorded into the base definition and does not infleucne the final
results.
\end{thebibliography}
\end{document}